\documentclass[twocolumn]{aastex63}

\usepackage{amsmath,amstext}
\usepackage[figure,figure*]{hypcap}

\received{receipt date}
\revised{revision date}
\accepted{acceptance date}
\published{published date}
\submitjournal{AASJournal name}

\shorttitle{Cold Co-condensation of Silicate and Carbon}
\shortauthors{Rouill\'e et al.}

\begin{document}

\title{Separate Silicate and Carbonaceous Solids Formed from Mixed Atomic and Molecular Species Diffusing in Neon Ice}
\author{Ga\"el Rouill\'e}
\affiliation{Laboratory Astrophysics Group of the Max Planck Institute for Astronomy at the Friedrich Schiller University Jena, Institute of Solid State Physics, Helmholtzweg 3, D-07743 Jena, Germany}
\author{Cornelia J\"ager}
\affiliation{Laboratory Astrophysics Group of the Max Planck Institute for Astronomy at the Friedrich Schiller University Jena, Institute of Solid State Physics, Helmholtzweg 3, D-07743 Jena, Germany}
\author{Thomas Henning}
\affiliation{Max Planck Institute for Astronomy, K{\"o}nigstuhl 17, D-69117 Heidelberg, Germany}

\correspondingauthor{Cornelia J\"ager}
\email{cornelia.jaeger@uni-jena.de}

\begin{abstract}
The formation and growth of refractory matter on pre-existing interstellar dust grain surfaces was studied experimentally by annealing neon-ice matrices in which potential precursors of silicate grains (Mg and Fe atoms, SiO and SiO$_2$ molecules) and of solid carbon (C$_n$ molecules, $n$ = 2--10) were initially isolated. Other molecules, mainly O$_3$, CO, CO$_2$, C$_3$O, and H$_2$O, were embedded at the same time in the matrices. The annealing procedure caused the cold dopants to diffuse and interact in the neon ice. Monitoring the procedure in situ with infrared spectroscopy revealed the disappearance of the silicon oxide and carbon molecules at temperatures lower than 13~K, and the rise of the Si{\sbond}O stretching band of silicates. Ex situ electron microscopy confirmed the formation of silicate grains and showed that their structure was amorphous. It also showed that amorphous carbon matter was formed simultaneously next to the silicate grains, the two materials being chemically separated. The results of the experiments support the hypothesis that grains of complex silicates and of carbonaceous materials are re-formed in the cold ISM, as suggested by astronomical observations and evolution models of cosmic dust masses. Moreover, they show that the potential precursors of one material do not combine with those of the other at cryogenic temperatures, providing us with a clue as to the separation of silicates and carbon in interstellar grains.
\end{abstract}



\section{INTRODUCTION}

Solid particles of refractory substances are formed at high temperature in the outer shells and in the cooling outflows of evolved stars \citep[e.g.,][]{Ferrarotti06}, and in supernova ejecta \citep[e.g.,][]{Sugerman06,Matsuura11}. They are eventually injected into the interstellar medium (ISM) to evolve as interstellar dust grains. Various mechanisms alter and eventually destroy them over a time-scale that estimations find shorter than the time-scale of their injection by stars \citep{Draine79,Jones96,Jones05,Draine09}. Astronomical observations, however, show that the mass of interstellar dust is steady. A mechanism that creates dust mass in the ISM would conciliate the two results \citep{Jones96,Jones05,Draine09,Jones11,Jones14}.

The mass of interstellar dust can increase locally through the accretion of atoms and molecules present in the interstellar gas phase, which necessitates the adsorption of these gas-phase precursors by existing particles. Moreover, the adsorption lifetime must make possible the chemical reactions that produce the solid material of interstellar dust. This suggests that the local creation of interstellar dust mass must take place in cold regions. Such regions include the cold neutral ISM, or cold H~{\small I} medium, where the temperature is $\sim$100~K, dense molecular clouds where the temperatures range from 10 to 50~K, and all regions with intermediate density and temperature conditions \citep{Draine11}. It is however not understood how the refractory components of the interstellar dust can be produced at these cryogenic temperatures. Additionally, the role of the interstellar radiation field in the diffuse regions and that of ice mantles in dense ones have to be clarified.

Interstellar dust comprises essentially silicate grains and carbonaceous particles. Experimental studies reporting the formation of these refractory substances at low temperatures are scarce. Silicate-related solids \citep{Donn81,Khanna81} and carbonaceous particles \citep{Wakabayashi04,Wakabayashi05} were obtained following the annealing of ice matrices in which specific atomic and molecular species were initially isolated. In another study, a silicate residue was formed by proton irradiation and warming of H$_2$O ices containing SiH$_4$ molecules \citep{Nuth89}. More recently, experiments by our group have demonstrated the formation of complex silicates \citep{Rouille14} and of carbonaceous solid matter \citep{Fulvio17} at temperatures not higher than 13 and 15~K, respectively. In liquid helium, at 1.7~K, carbon atoms and molecules were found to condense into a partially graphitized solid \citep{Krasnokutski17}. In such conditions, a chemistry that requires very little activation energy, if any, is involved \citep{Krasnokutski14c}.

These observations made in the laboratory support the hypothesis that low temperature is not an obstacle for the formation and the growth of refractory grains in the cold regions of the ISM. Another issue is the separation of interstellar dust into carbonaceous particles and silicate grains suggested by some studies \citep{Adamson99,Dwek04,Chiar06,Smith06,Mason07,Li14,Valencic15}. The formation and the growth of mostly pure grains in a medium populated with the atomic and molecular precursors of both silicate and carbonaceous materials, not mentioning other substances, would necessitate selective adsorption and desorption mechanisms.

Studying the co-accretion of cold precursors of silicates and carbonaceous solids, together with other species relevant to the ISM, would contribute to explaining the separation of the two refractory materials in the interstellar dust. It would also give us the opportunity to demonstrate that silicates and carbonaceous solids can be formed and can grow even though other species are present. Finally, it is expected to show that solid SiC is not formed in interstellar conditions, as can be inferred from its low abundance in the diffuse ISM \citep{Whittet90,Chiar06a}. Accordingly, we present an experimental study examining the simultaneous accretion of potential precursors of silicate grains and solid carbon at cryogenic temperatures, in the presence of species relevant to the ISM.

\section{EXPERIMENTAL}\label{sec:exp}

\begin{figure*}
\epsscale{1.1}
\plotone{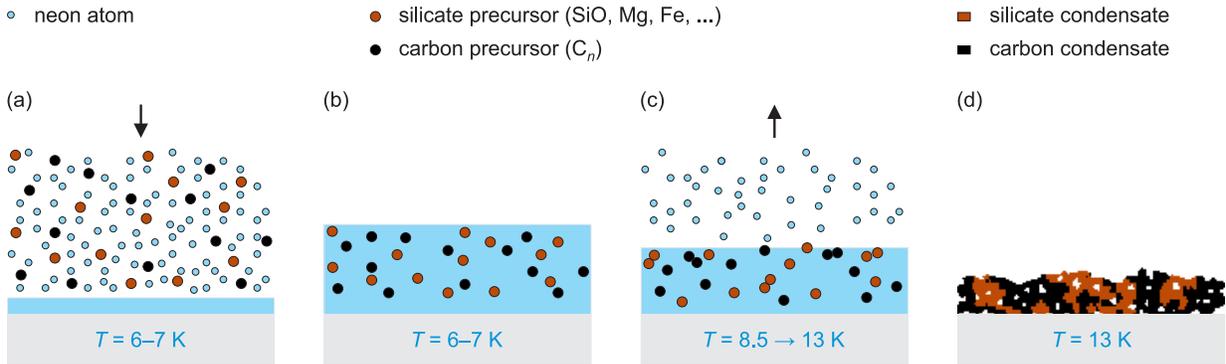}
\caption{Principle of the experimental procedure used to study the simultaneous condensation of silicates and carbonaceous grains. (a) Deposition of atomic and molecular precursors produced by laser vaporization, with Ne atoms, on a cold KBr substrate. (b) Cold precursors isolated in Ne ice and identified with absorption spectroscopy. (c) Annealing of the Ne ice causing the diffusion and accretion of the cold precursors, monitored with FTIR spectroscopy. (d) At 13~K, atomic and molecular precursors have disappeared and a refractory condensate is observed. Impurities (e.g., H$_2$O, CO, CO$_2$) that form ices are not included.\label{fig:process}}
\end{figure*}

\begin{figure}
\epsscale{1.1}
\plotone{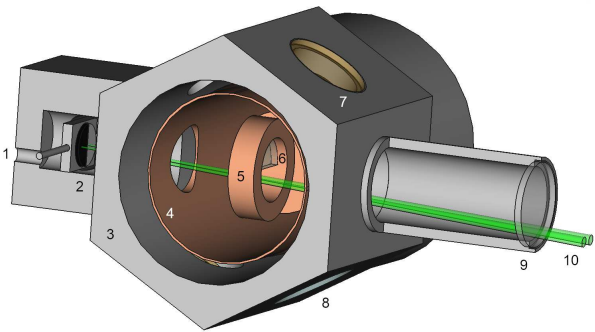}
\caption{Core of the apparatus used for the experiments. (1) Ne gas inlet; (2) vacuum chamber extension with targets; (3) vacuum chamber, with evacuation (not shown) toward the foreground; (4) radiation shield fixed to the first stage of the rotatable cryocooler; (5) substrate holder fixed to the second stage; (6) semicircular substrate; (7) port for IR spectroscopy; (8) port for UV/vis spectroscopy; (9) extension with port for laser vaporization; (10) laser beams. The front of parts (2), (3), (4), and (9) has been cut away.\label{fig:setup}}
\end{figure}

The principle of the experiments is to make cryogenic-cold atomic and molecular species interact with each other in conditions that enable their accretion into a solid, refractory material. Figure~\ref{fig:process} illustrates the procedure we have applied and Figure~\ref{fig:setup} shows the core part of the apparatus we have used.

As the first step of several experiments, the potential atomic and molecular precursors of silicates and of carbonaceous matter were produced by the simultaneous laser vaporization of a silicate-related target and a graphite target (Plano GmbH). They were placed side by side in a vacuum chamber with a base pressure of $\sim$1$\times$10$^{-6}$~mbar at room temperature and $\sim$1$\times$10$^{-7}$~mbar at 6.5~K. Four experiments denoted E1, E2, E3, and E4 were carried out, each time with a different silicate-related target, so as to observe the influence of various metal atom concentrations. To date, interstellar silicates are found to be Mg-rich \citep{Min07,Fogerty16}, a large fraction of the depleted interstellar Fe atoms being possibly found as metallic Fe and FeS nanometer-sized inclusions \citep{Koehler14,Westphal14}. In experiments E1 to E4, the targets were respectively pressed SiO powder (Sigma Aldrich) and slabs of quenched melts with the formulas Mg$_2$SiO$_4$, Mg$_{0.4}$Fe$_{0.6}$SiO$_3$, and Mg$_{0.6}$Fe$_{0.4}$SiO$_3$. The three compounds with complex silicate compositions were amorphous, glassy. They were synthesized in-house following the melting procedure described by \cite{Jaeger94} and \cite{Dorschner95}.

Cooling of the atomic and molecular products of laser vaporization down to $\sim$7~K was achieved by isolating them in Ne ice. For this purpose, the targets to be vaporized were placed 5.5~cm away from a KBr substrate (Korth Kristalle GmbH) kept at $\sim$6.5~K by the action of a compressed-He, closed-cycle cryocooler (Advanced Research Systems, Inc. ARS-4H and DE-204SL). Laser vaporization was performed during 60 to 65~min with synchronized pulsed Nd:YAG sources (Continuum Minilite I and Minilite II) emitting at 532-nm wavelength and operated at a rate of 10 shots per second. The laser spots on the targets were shifted every minute. At the same time, a continuous flow of Ne gas (Air Liquide, purity 99.999{\%}) was directed towards the cold substrate with a rate of 5.00~sccm, bringing the pressure to $\sim$2$\times$10$^{-5}$~mbar in the chamber. In such conditions the substrate became covered with Ne ice in which the products of laser vaporization were embedded.

The second step was the annealing of the Ne ice to cause the diffusion and the interaction of the cold atoms and molecules it contained. Any energy produced by chemical reactions would be dissipated into the ice, preventing the dissociation of the products and enabling accretion. The annealing consisted of increasing the temperature of the KBr substrate by electrical heating (temperature controller LakeShore 330 Autotuning). The increase was carried out at an overall rate of $\sim$0.02~K~min$^{-1}$, starting from 8.5--9.0~K (with oscillations of $\pm$0.5~K). The Ne ice disappeared completely near 13~K as Ne ice evaporates at an extremely fast rate at this temperature \citep[see, e.g.,][]{Hama17}.

In situ optical spectroscopy at ultraviolet and visible (UV/vis) wavelengths and also in the infrared (IR) domain allowed us to verify the presence of the precursors isolated in the Ne ice prior to annealing. It also allowed us to monitor the disappearance of these species during the annealing procedure, and, possibly, the associated formation of a solid material at low temperature.

Finally, the solids, or condensates in the broad sense, produced in the experiments were also studied ex situ by high-resolution transmission electron microscopy (HRTEM) and energy-dispersive X-ray (EDX) spectroscopy. These techniques provided us with information on the morphology and the composition of the condensates, respectively. In order to perform their HRTEM analysis, the solids produced in the experiments were transferred from the KBr substrate to a TEM copper grid covered with a lacey carbon film. The HRTEM analysis was performed with a transmission electron microscope (JEOL GmbH JEM-3010) equipped with a LaB$_6$ cathode and operated with an acceleration voltage of 300~kV. The HRTEM micrographs were Fourier-transformed (FT) to reveal periodic structures such as lattice fringes or structural patterns related to a medium range order of the amorphous material. Non-physically related periodicities were removed from the inverse-FT images and subsequent reconversion into bright-field images allowed us to produce HRTEM micrographs of high quality. Some were skeletonized to illustrate differences in the medium-range order, which indicates the degree of structural ordering in amorphous materials. For instance, in carbonaceous grains, the presence of plane or bent aromatic structural units is described by the medium-range order. In contrast, a crystalline structure is a display of long-range order.

\section{RESULTS}

\subsection{Identification of the Dopants}

\begin{figure}
\epsscale{1.1}
\plotone{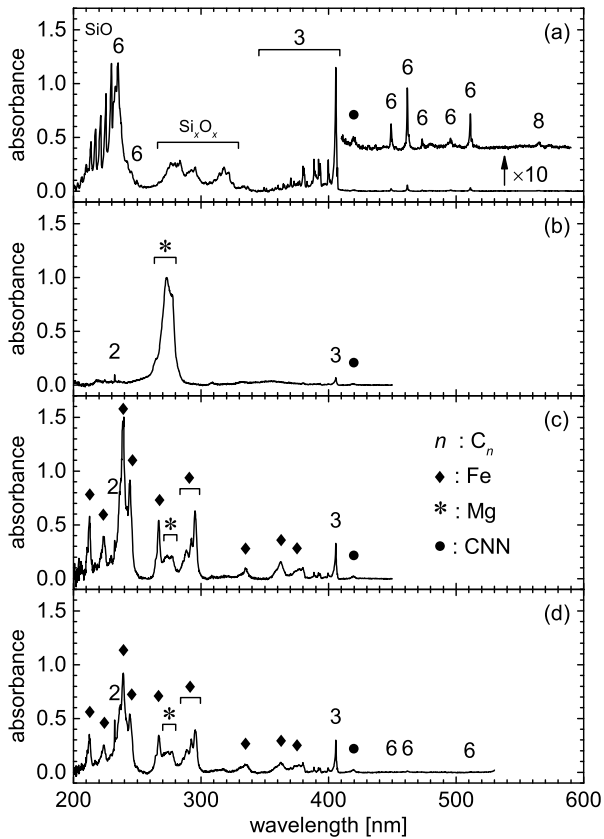}
\caption{Absorption spectra of doped Ne matrices in the UV/vis wavelength domain after subtraction of the baselines. Matrices obtained after (a) 64, (b) 10, (c) 12, and (d) 11~min deposition of Ne atoms while carrying out laser vaporization of a graphite target paired with a target of (a) SiO, (b) Mg$_2$SiO$_4$, (c) Mg$_{0.4}$Fe$_{0.6}$SiO$_3$, and (d) Mg$_{0.6}$Fe$_{0.4}$SiO$_3$. As to peak labels, the numbers $n$ refer to C$_n$ molecules, the black diamonds to atomic Fe, the asterisks to atomic Mg, and the black bullets to CNN.\label{fig:UVvis}}
\end{figure}

Figure~\ref{fig:UVvis} shows UV/vis spectra of Ne matrices doped with atoms and molecules produced in the laser vaporization of the silicate-related and graphite targets. The species identified by examining the UV/vis spectra are Mg, Fe, C$_2$, C$_3$, C$_6$, C$_8$, SiO, and CNN. Table~\ref{tbl:UVvisspecies} contains information on the wavelength positions of the features used for identification. Not included in Table~\ref{tbl:UVvisspecies} are complex absorption features attributed to Si$_x$O$_x$ ($x$ $>$ 1) oligomers \citep{Rouille13,Krasnokutski14c}.

When present, Mg and Fe atoms were identified by using reference spectra of these atoms isolated in Ne ice. The atoms were then produced by laser vaporization of pure metal targets. There are no clear signs of the presence of Mg or Fe dimers in the present spectra. Interestingly, Si atoms do not appear in any of the matrices. We infer that the Si atoms are all contained in the various silicon oxide molecules observed in our experiments, because there are no features that could be attributed to other silicon-containing molecules, e.g., silicon carbides.

The $A ^1\Pi \leftarrow X ^1\Sigma^+$ transition of SiO gives lines in the UV/vis wavelength range. It is visible in the spectrum obtained with the SiO target because it was measured after a long deposition time (64~min). The spectra obtained with the silicate targets were measured after depositing materials for 10--12~min only, so as to avoid saturation by the strongly absorbing atomic lines.

The detection of CNN in the matrices indicates the production of C atoms via the laser vaporization of graphite. The observed CNN molecules are the product of a barrierless reaction between C atoms and N$_2$ molecules of the background gas \citep{Hickson16}. The dissociation of the product was prevented by the dissipation of the reaction energy by the rare-gas ice. The direct detection of C atoms in the Ne matrices was not achieved due to the lack of practical features in the wavelength domains scanned in the experiments.

As said above, the detection of CNN molecules indicates the presence of a significant amount of background gas in the vacuum chamber. The molecules N$_2$, O$_2$, CO, CO$_2$, and H$_2$O were likely embedded in the rare-gas ice. Accordingly, the features of CO, CO$_2$, and H$_2$O appeared in the IR spectra.

\begin{figure*}
\epsscale{1.1}
\plotone{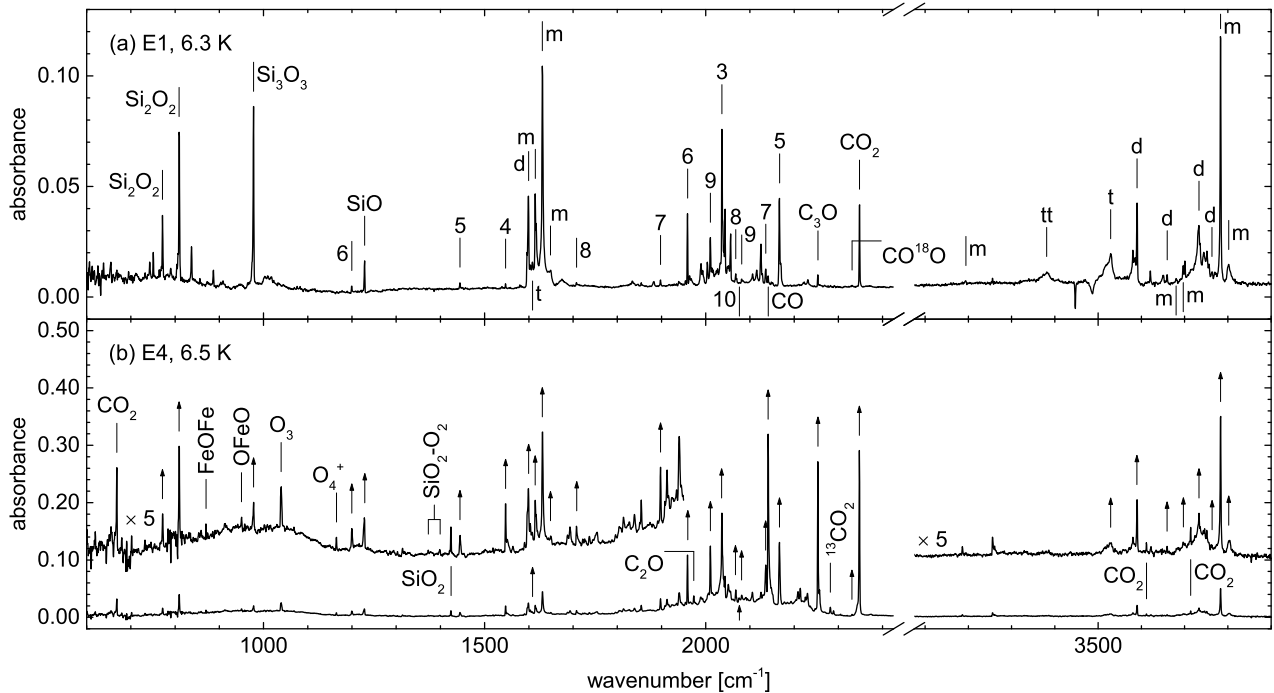}
\caption{Infrared spectra of Ne matrices doped with atoms and molecules produced in the laser vaporization of two targets. (a) Experiment E1, graphite and SiO. (b) E4, graphite and Mg$_{0.6}$Fe$_{0.4}$SiO$_3$. In both panel, the interference pattern generated by the matrix and the general baseline have been subtracted. The numbers $n$ refer to C$_n$ molecules, the letters m, d, t, and tt stand for water monomer, dimer, trimer, and tetramer. In panel (b), arrows indicate bands identified in panel (a) and five-time magnified sections of the spectrum are vertically offset for clarity.\label{fig:MIR1}}
\end{figure*}

\begin{figure*}
\epsscale{1.0}
\plotone{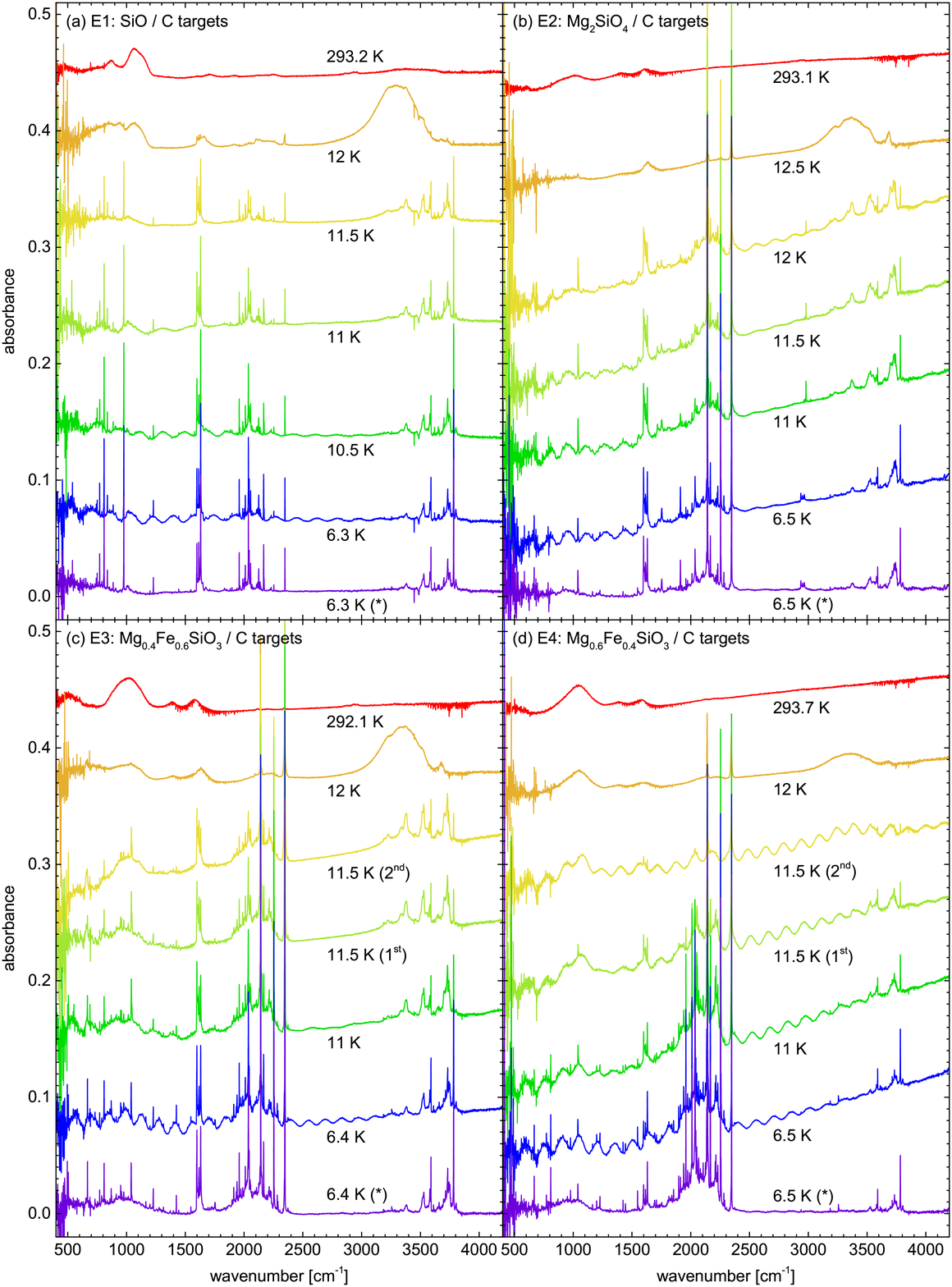}
\caption{Infrared spectra measured before and during the annealing of Ne matrices doped with atoms and molecules produced in the laser vaporization of two targets. (a) Experiment E1: graphite and SiO. (b) E2: graphite and Mg$_2$SiO$_4$. (c) E3: graphite and Mg$_{0.4}$Fe$_{0.6}$SiO$_3$. (d) E4: graphite and Mg$_{0.6}$Fe$_{0.4}$SiO$_3$. The spectra marked with (*) are those measured at 6.3-6.5~K after subtracting the interference pattern generated by the matrix, and also the general baseline. Labels 1$^{\mathrm{st}}$ and 2$^{\mathrm{nd}}$ indicate measurements made at close times. The spectra are offset vertically for clarity.\label{fig:MIR2}}
\end{figure*}

\begin{figure}
\epsscale{1.1}
\plotone{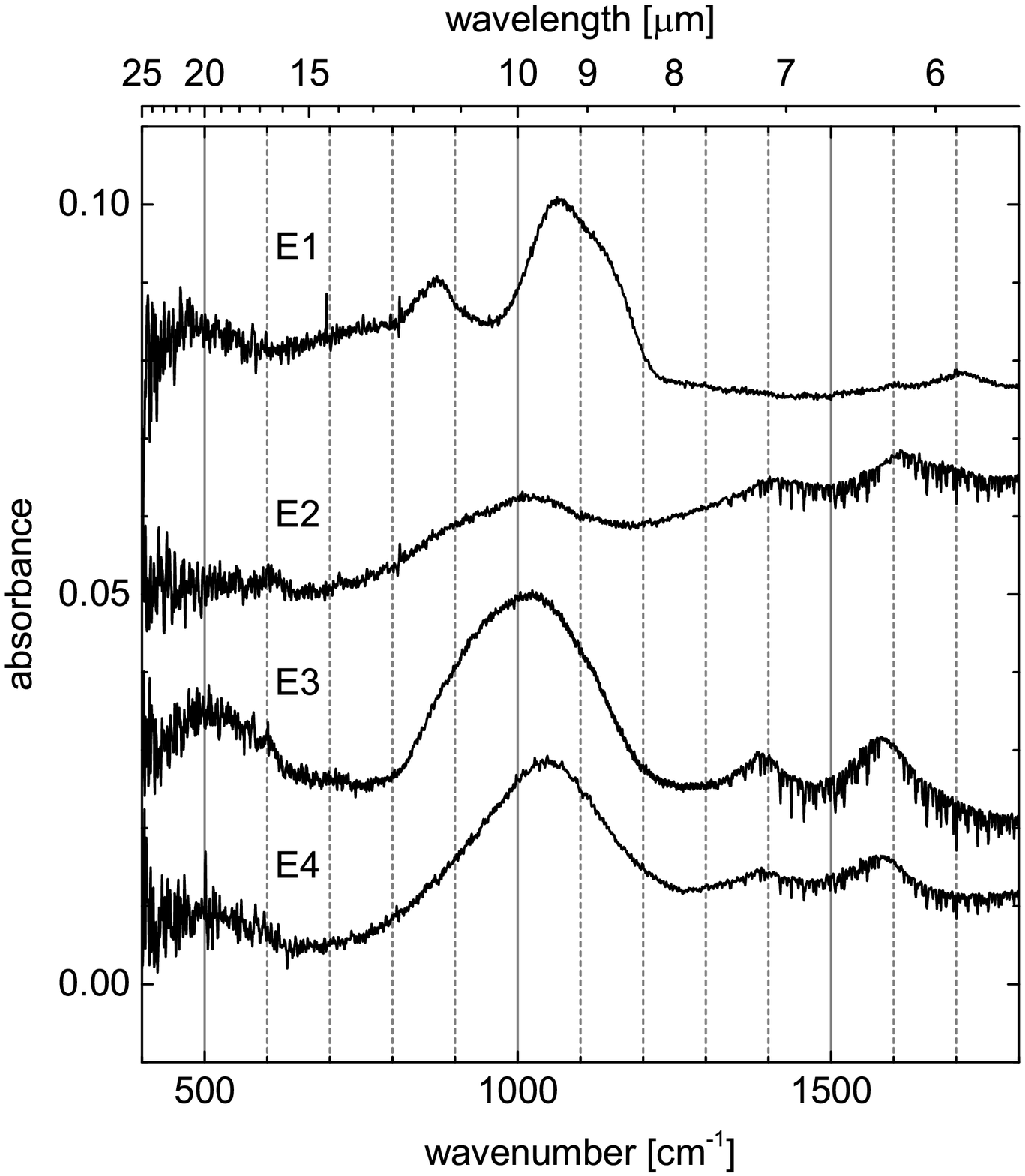}
\caption{Infrared spectra of the materials formed in the experiments E1 to E4, measured at room temperature. Lines of water vapor are not fully corrected because of varying vacuum conditions in the spectrometer during experiment. The spectra are offset vertically for clarity.\label{fig:MIR3}}
\end{figure}

Figures~\ref{fig:MIR1} to~\ref{fig:MIR3} present Fourier-transform IR (FTIR) spectra measured during the experiments. The list of the matrix-isolated species detected and identified by analyzing these spectra, the frequencies of the absorption bands attributed to the species, and literature references are given in Table~\ref{tbl:obsvib}.

The assignment of a band to Ne-matrix-isolated SiO$_2$ was tentative in our earlier work on the formation of silicates \citep{Rouille14}. It was then suggested by studies reporting the band of Ar-matrix-isolated SiO$_2$ at 1416.4~cm$^{-1}$ \citep{Andrews92,Tremblay96}. The assignment was confirmed when we performed the laser vaporization of an amorphous Mg$_{0.4}$Fe$_{0.6}$SiO$_3$ target to dope an Ar matrix and found at 1417.0~cm$^{-1}$ the band seen at 1423.8~cm$^{-1}$ in Ne matrices.

The absorption bands of the carbon oxide molecules CO, CO$_2$, and C$_3$O are also detected. In the case of C$_3$O, the identification relies on spectra of the molecule in the gas phase, isolated in Ar ice, and isolated in Ne matrix (D. Strelnikov, private communication). The  amount of carbon oxide molecules isolated in the matrices is much larger when working with a silicate target than with an SiO target. In the former case, the bands of CO, CO$_2$, and C$_3$O (at 2141.2, 2347.5, and 2253.5~cm$^{-1}$, resp.) dominate the FTIR spectrum and show a different pattern of relative intensities. This suggests that the laser vaporization of a silicate target, as performed here, produces oxygen species that react with carbon species generated in the simultaneous laser vaporization of the graphite target. The reaction may take place in the gas phase or at and under the surface of the growing Ne matrix where the atoms and/or molecules may diffuse. By contrast, the vaporization of SiO essentially produces SiO molecules and their oligomers. In that case, the carbon oxide molecules are either background molecules or the products of reactions between the carbon species produced by laser vaporization and the oxygen molecules of the background gas phase.

Similarly, the oxygen molecules O$_3$ and O$_4^+$ were detected only in the experiments with silicate targets. It can be assumed that they are not produced when shooting at a SiO target because this material is vaporized essentially in the form of SiO molecules.

Water molecules were initially present in the Ne matrices. They came from the background gas in the vacuum chamber and likely from the targets as well, which were not treated before the experiments to remove adsorbed water. The amount of H$_2$O molecules in the doped Ne matrices was large, causing the formation of dimers, trimers, tetramers, and some pentamers. It is indicated by the broad features arising between 3300 and 3800~cm$^{-1}$.

An unknown species that gives bands at 913 and 2192~cm$^{-1}$ was detected in the experiments with laser vaporization of silicate targets. It was already spotted in our previous study on silicates \citep{Rouille14}. This species was reported by \cite{Jacox13} who proposed a complex involving H$_2$ and H$_3$O$^+$ or H$_2$O$_5^+$ as a possible carrier. Since it is not identified, we neglect its possible role in the formation of refractory materials.

Unassigned weak absorption bands indicate the presence of other species beside the main substances listed above. They likely comprise van der Waals complexes formed with the most abundant dopants, and various oxides. For instance, a band is tentatively attributed to CO$\sbond$H$_2$O, and one of two features between 2050 and 2057~cm$^{-1}$ may be caused by C$_3$$\sbond$H$_2$O \citep[][in Ar matrix]{Szczepanski95,Dibben00}. A band seen at $\sim$1934~cm$^{-1}$ when using the iron-containing targets may signal the formation of FeCO \citep{Zhou99} already during the growth of the matrices.

We did not find any absorption that could be attributed to molecular MgO, which has a vibrational frequency of $\sim$775~cm$^{-1}$ in the gas phase \citep{Kagi06}.

\subsection{Amounts of Dopants}

The amount of a dopant in a matrix can be evaluated from its absorbance spectrum provided that the thickness $d$ of the matrix and the molecular absorption cross-section of the dopant are known. We have neglected the effect of the matrix on the absorption cross-section of the free species. Because the refractive index $n$ is not known, only the optical equivalent thickness $nd$ was determined and used in place of $d$. It was estimated by analyzing the interference patterns observed in the MIR spectra. Thus $nd$ was determined to be 28.6, 31.3, 35.3, and 33.3~$\mu$m in experiments E1, E2, E3, and E4, respectively. The areas of the atomic lines identified in the UV/vis spectra were compared with the cross-sections of the corresponding transitions. The cross-sections were calculated by using the vacuum wavelengths, Einstein coefficients, and state degeneracies \citep{NIST_ASD}.

Table~\ref{tbl:concentrations} gives the amount of the identified and most significant dopants of the Ne matrices in the various experiments.

\begin{deluxetable}{lllll}
\tablecaption{Average Concentrations\tablenotemark{a} of Matrix-isolated Species\label{tbl:concentrations}}
\tablehead{\colhead{Species} & \colhead{E1\tablenotemark{b}} & \colhead{E2} & \colhead{E3} & \colhead{E4} }
\startdata
Mg                      &       & 10.06  & 1.55  & 1.64 \\
Fe                      &       &        & 19.2  & 16.2 \\
SiO                     & 26.0  & 19.6   & 59.3  & 41.1 \\
SiO$_2$                 &       & 6.63   & 17.9  & 11.3 \\
Si$_2$O$_2$             & 16.4  &        & 8.81  & 15.9 \\
Si$_3$O$_3$             & 6.76  &        &       & 1.14 \\
C$_2$                   &       & 1.53   & 2.83  & 4.99 \\
C$_3$                   & 10.0  & 4.90   & 11.7  & 35.2 \\
C$_4$                   & 0.538 & 2.41   & 2.17  & 4.58 \\
C$_5$                   & 2.02  & 3.41   & 3.53  & 8.49 \\
C$_6$                   & 2.55  & 1.15   & 5.15  & 9.91 \\
C$_7$                   & 0.297 & 1.72   & 0.551 & 1.00 \\
C$_8$                   & 0.243 & 0.0972 & 0.390 &      \\
C$_9$                   & 0.498 & 0.156  & 0.554 & 1.55 \\
CO                      & 8.92  & 610    & 477   & 458  \\
CO$_2$                  & 5.85  & 70.4   & 65.8  & 51.5 \\
C$_3$O                  & 0.330 & 14.5   & 13.2  & 19.0 \\
H$_2$O\tablenotemark{c} & 425   & 213    & 483   & 180  \\
O$_3$                   &       & 24.9   & 30.2  & 13.9 \\
\enddata
\tablenotetext{a}{In units of 10$^5$~$\mu$m$^{-3}$.}
\tablenotetext{b}{Experiments E1, E2, E3, and E4 used targets of SiO, Mg$_2$SiO$_4$, Mg$_{0.4}$Fe$_{0.6}$SiO$_3$, and Mg$_{0.6}$Fe$_{0.4}$SiO$_3$, respectively, along graphite.}
\tablenotetext{c}{The contributions of water clusters and water-containing van der Waals complexes are not included.}
\end{deluxetable}

When Fe and Mg atoms are simultaneously present in the Ne matrices, seven Lorentz profiles were fitted to the features observed in the 32\,701--38\,941~cm$^{-1}$ interval of energy (256.8--305.8~nm wavelength interval). The two profiles centered between 35\,000 and 38\,000~cm$^{-1}$ were assigned together to the single Mg~{\footnotesize I} line at 35\,051.253~cm$^{-1}$ in vacuum \citep{NIST_ASD}. The other features were assigned to Fe~{\footnotesize I} lines, given at 33\,095.9408, 33\,507.1232, 33\,695.3972, 34\,039.5154, and 36\,766.9660~cm$^{-1}$, in vacuum \citep{NIST_ASD}. When only Mg atoms were present, four Lorentz profiles were fitted to the double-peaked feature and its base between 33\,112 and 41\,667~cm$^{-1}$. Only the two main Lorentz profiles were taken into account to determine the amount of Mg atoms in the matrix.

Regarding C$_2$, the origin band of the Mulliken $D ^1\Sigma_u^+$--$X ^1\Sigma_g^+$ system was found at 232.2~nm. It was used to determine the column density of C$_2$ by giving it an $f$-value or oscillator strength of 0.0545 for a central wavelength of 231.3~nm in air \citep{Lambert95}. The integrated molecular absorption cross section for this transition was determined using these values of strength and wavelength \citep{Mulliken39}. Its comparison with the integrated absorbance of the band  observed in our spectra yields allowed us to derive the density of C$_2$ molecules in the Ne matrices.

In the case of infrared-active molecules, bands observed in the FTIR spectra were exploited to determine the amount of the molecular dopants in the Ne matrices. The amount was derived from the comparison between measured band areas and vibrational intensities computed at the density functional theory (DFT) level with the Gaussian 03 program \citep{Gaussian03}. We used the B3LYP functional \citep{Becke88,Lee88,Becke93} in combination with the 6-311+G(d,p) basis set \citep{Krishnan80,McLean80,Frisch84}.

The data in Table~\ref{tbl:concentrations} can be used to evaluate the number of atoms that can contribute to forming the solid condensates. In particular, Mg and Fe atoms, Si atoms from oxides, and C atoms from carbon molecules. Most C atoms in carbon oxide molecules would likely escape as the molecules get into the gas phase. Thus, Si and C atoms are present in numbers of the same order. The number of Fe atoms is at least five times smaller in comparison, though those deposited as oxides are not being counted. They are, however, potential precursors of silicates. The number of Mg atoms is an order of magnitude smaller that that of Fe atoms in the experiments E3 and E4. Magnesium oxides are not taken into account because they do not appear in the FTIR spectra.

The contents of Table~\ref{tbl:concentrations} show that the Ne matrices in experiments E2, E3, and E4 with complex silicate targets were rich in CO and H$_2$O molecules in comparison with silicon oxide molecules and carbon molecules. This did not prevent the formation of silicate and carbonaceous condensates.

\subsection{Condensation Processes}

As the Ne ice was annealed, the various species it contained diffused and interacted, thus giving barrierless chemical reactions the opportunity to take place. Reactions were possible between atoms and molecules and between an atom or molecule and a site at the surface of a cluster or a particle.

During the processing of the rare-gas matrix, the number of isolated water molecules decreases as they form dimers, etc. and water ice is formed as seen in Figure~\ref{fig:MIR2}. The features of water ice are visible after the Ne atoms have disappeared. They are broad and found in the IR spectra at 770, 1650, and 3280~cm$^{-1}$, or 13.0, 6.06, and 3.05~$\mu$m, respectively \citep[see, for instance,][]{Oeberg07}. Water ice sublimates when the substrate reaches a temperature of $\sim$160~K under vacuum \citep[value for pure H$_2$O ice,][]{Collings04} and its features are no longer visible in the spectra measured at room temperature.

Some of the water molecules can react chemically with other species or with surface groups of particles such as Si$\sbond$O$\sbond$Si or aromatic C$\sbond$H. These reactions, however, have not been studied at low temperatures yet. On the other hand, reactions of H$_2$O with C and with Fe were studied at low temperature in He droplets and in Ar ice. They were found to give the weakly bond compounds H$_2$O$\sbond$C \citep{Krasnokutski14a} and H$_2$O$\sbond$Fe \citep{Krasnokutski14b,Deguin18}.

The oxygen molecules O$_2$ and O$_3$ can contribute to forming ice and can also react chemically with other species to produce potential silicate precursors. For instance, Mg atoms can react with O$_2$ \citep{Krasnokutski10} and with O$_3$ \citep{Andrews78} to give magnesium oxides. Additionally, Fe atoms can react with O$_2$ to form the weakly bound FeOO molecule \citep{Krasnokutski14b}.

In Ne matrices, CO molecules can react with Fe atoms to produce Fe$_x$(CO)$_y$ compounds \citep{Zhou99}. The most abundant product would be FeCO with a band at 1933.7~cm$^{-1}$. A weak band visible in the spectra obtained when using the Fe-containing targets coincides with this absorption. Its assignment is not certain.

Nitrogen molecules contribute to ice formation during the annealing of the Ne matrix. They would start to sublimate around 20~K \citep{Collings04}. The complex composition of the ice (H$_2$O, N$_2$, carbon oxides, O$_2$, etc.) may make the sublimation irregular.

During the annealing of the Ne ice, SiO and its oligomers react together despite the low temperature as the reactions do not require activation energy \citep[][and references therein]{Krasnokutski14c}. These reactions alone produce silicon oxide solids that give rise to the 10- and 20-$\mu$m bands in Figure~\ref{fig:MIR2}(a) \citep{Rouille13,Krasnokutski14c}. In addition, the silicon-bearing molecules react with Mg, Fe, and their oxides to form complex silicates such as MgSiO$_3$ and Mg$_{0.24}$Fe$_{0.76}$SiO$_3$ \citep{Rouille14}, and the broad band that arises at 10-$\mu$m in Figures~\ref{fig:MIR2}(b) to \ref{fig:MIR2}(d) is attributed to such solid compounds. It is assumed that the SiO$_2$ molecules contribute to forming these condensates as well.

The bands of the matrix-isolated iron oxide molecules, which are detected in experiments with Fe-containing silicate targets, disappear during the heating and sublimation of the Ne ice. It is assumed that they are incorporated in the forming refractory condensates as mentioned above.

Chemical reactions at cryogenic temperatures between carbon molecules were proposed to explain the evolution of IR bands during the annealing of rare-gas matrices doped with such species \citep{Thompson71}. Experiments \citep{Wakabayashi04} and molecular dynamics modeling \citep{Yamaguchi04} suggested that condensates of amorphous carbon can be formed in a cold environment \citep{Wakabayashi05}. The phenomenon was demonstrated experimentally by \cite{Fulvio17}. Presently we observe the attenuation and disappearance of the bands caused by C$_n$ molecules as the Ne ice is warmed up and evaporated. While the formation of a refractory carbonaceous material cannot be monitored by the appearance of a strong, distinct IR band, it is nonetheless related to a general increase of the absorption due to the formation of free charge carriers. An increase of the absorption was observed during annealing even though it had to be distinguished from occasional variations of the baseline vertical position. Electron microscopy shows that solid amorphous carbon was formed (see Section~\ref{sec:TEM}).

Two broad bands are found in the IR spectra of the condensates produced in the experiments with the silicate and graphite targets. Their wavelength positions change slightly with the Fe-content of the silicate: 6.2 and 7.1~$\mu$m (resp. 1610 and 1410~cm$^{-1}$) with the Fe-free silicate target, 6.3 and 7.2~$\mu$m (resp. 1583 and 1392~cm$^{-1}$) with the two Fe-containing silicate targets. Because the bands do not appear in experiments without carbon \citep{Rouille14} or without magnesium, they are attributed to magnesium and magnesium-iron carbonates. The degenerate asymmetric stretching mode $\nu_3$ of the CO$_3^{2-}$ ion, which exhibits the symmetry elements of the $D_{3h}$ point group, gives rise to a band near 7~$\mu$m. The band is split into the two components presently observed when the ion is distorted by its environment \citep{Brooker01}. Our assignment is consistent with the high abundance of carbon oxide and water molecules in the Ne matrices. Note that the spectra of the freshly prepared matrices did not show the features of isolated CO$_3^{2-}$ \citep[][in Ar matrix]{Jacox74}.

Two features that resemble the 6.3- and 7.2-$\mu$m bands discussed above appear in the IR spectrum of the condensate produced with the SiO and graphite targets. They differ in position, however, as they peak at 5.85 and 6.25~$\mu$m (1709 and 1600~cm$^{-1}$, respectively). In this condensate, the bands arise likely from water molecules bound in two ways, with and without hydrogen bonding \citep[e.g.,][]{Frost09a}. This would be consistent with the broad band at $\sim$3400~cm$^{-1}$.

Finally, the IR spectra show the formation of a broad band at 10~$\mu$m (1000~cm$^{-1}$) after annealing of the Ne matrix up to 12~K. This band is a clear signature of the low-temperature condensation of refractory silicate materials characterized by the typical Si$\sbond$O stretching band. In the presence of water, this band is partly obscured by the water libration band at 12.8~$\mu$m (781~cm$^{-1}$). The strength of the absorption at 10~$\mu$m increases as the Ne ice is annealed and eventually evaporated at $\sim$13~K, and the band appears clearly after water ice is removed by the final warming to room temperature.

\subsection{Composition and Structure of the Condensates}\label{sec:TEM}

\begin{figure*}
\epsscale{1.1}
\plotone{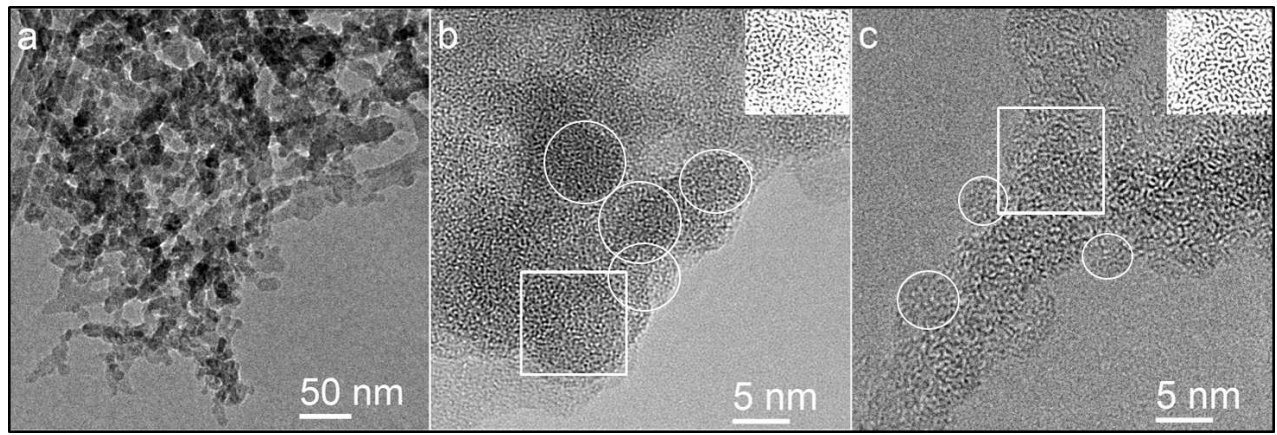}
\caption{(a) Representative TEM image showing the porous aggregate structure of mixed refractory condensates (E2). (b) and (c) are high-resolution micrographs of typical silicate and carbonaceous grains, respectively, observed in the condensate of experiment E2. Circles outline a few individual grains to demonstrate their sizes. The insets show noise-filtered images of the areas delimited with squares, illustrating differences in the medium-range order of both materials. The carbon grains are slightly smaller than the silicate particles and show a more distinct medium-range order characterized by small, bent graphene layers visible in edge-on view as black, curvy stripes of various lengths. They can be described as fullerene-like carbon grains. They can be described as fullerene-like carbon grains. Particle sizes, medium-range order, and EDX analysis were used to distinguish between carbonaceous and siliceous materials in the mixed condensates.\label{fig:HR1}}
\end{figure*}

\begin{figure*}
\epsscale{1.1}
\plotone{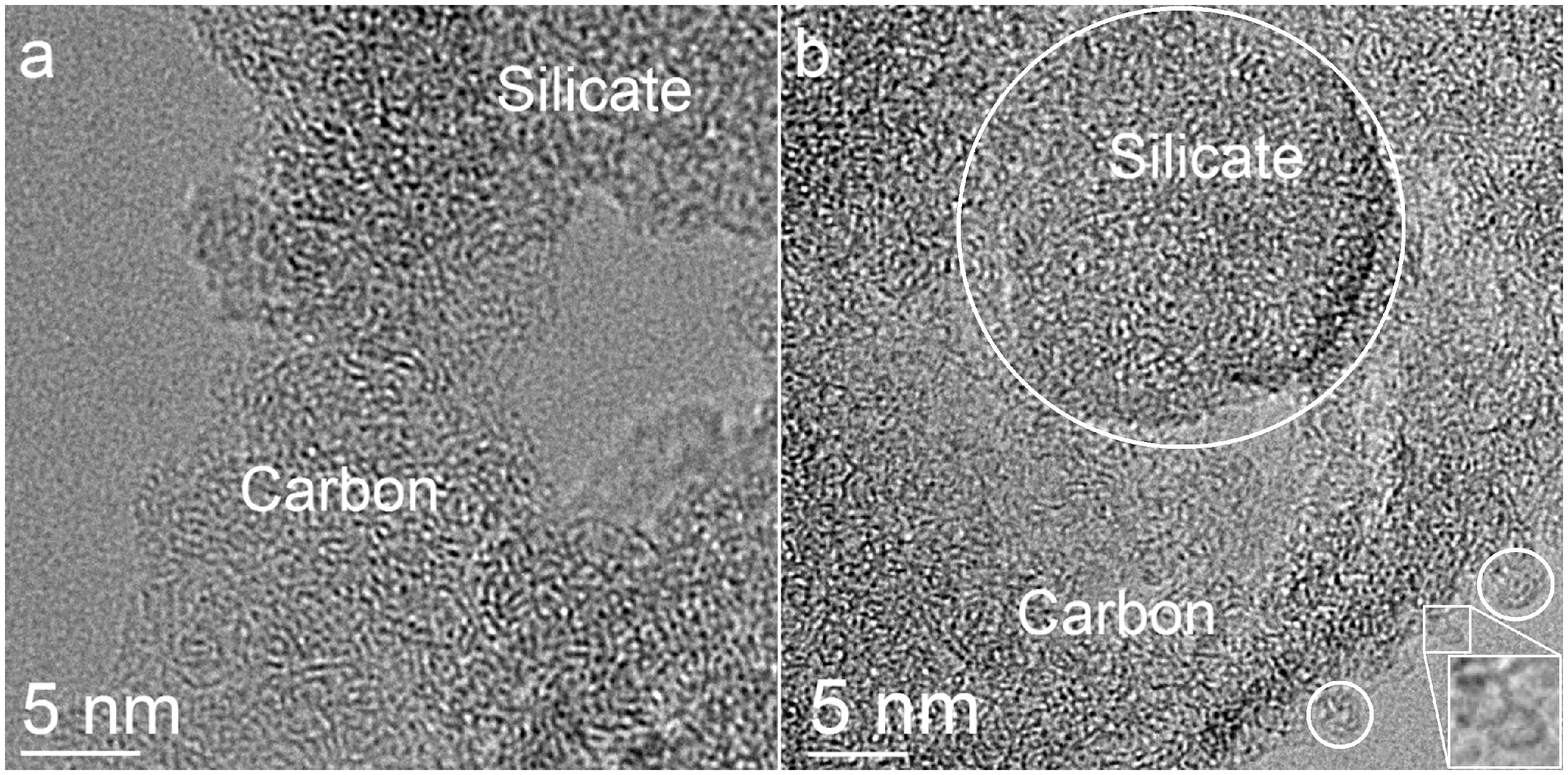}
\caption{HRTEM micrographs of two sample areas in the condensate of experiment E2, where silicate and carbonaceous grains are in contact. The medium-range order of the carbonaceous material characterized by small, bent graphene layers is illustrated in both images. In panel (b), a 25-nm-large agglomerate of three silicate particles, outlined with a circle, is surrounded and covered (on top) by carbon material, which consists of particles with sizes between 2 and 6~nm. Two carbon particles, marked with circles, can be more easily distinguished at the edge of the condensate. Two fullerene molecules at the periphery of a denser carbon cluster are framed in a square and magnified 2.5 times.\label{fig:HR2}}
\end{figure*}

Electron microscopy studies of the condensed refractory material verified the formation of porous aggregates of nanometer-sized particles as illustrated with Figure~\ref{fig:HR1}. The HRTEM analysis of the grains revealed the presence of chemically separated silicate and carbonaceous phases. Both materials are present as individual grains forming either pure or mixed aggregates, as shown in Figures~\ref{fig:HR1} and \ref{fig:HR2}, respectively. In Figure~\ref{fig:HR2}(c), a silicate particle is covered with carbonaceous material. The formation of SiC was not observed, neither with IR spectroscopy nor with HRTEM/EDX analysis. Both silicate and carbonaceous grains are characterized by an amorphous structure, which is clearly visible in Figures~\ref{fig:HR1}(b) and \ref{fig:HR1}(c). This common property makes it complex to distinguish between the two materials with HRTEM. Note that amorphous carbonates cannot be distinguished from amorphous silicates. For a distinct discrimination of the two condensates, we used EDX spectroscopy in combination with an image analysis that demonstrates the differences in the medium-range order of the amorphous materials. The medium-range order images were derived by processing the HRTEM images as described in Section~\ref{sec:exp}.

The micrographs of separated silicate and carbonaceous grains illustrate the smaller size of the primary carbon grains (2 to 6~nm) compared with the silicate particles (5 to 10~nm). As an additional difference, carbon grains show a distinct medium-range order characterized by strongly bent graphene layers, which can be directly identified as a substructure or can be derived from an image analysis. The slight dissimilarity between the carbonaceous and siliceous components is visible in the insets of Figures~\ref{fig:HR1}(b) and \ref{fig:HR1}(c). The structure of the carbonaceous material can be described as fullerene-like, and it is identical to that of the product in experiments on the condensation of pure carbon \citep{Fulvio17}. Like in the pure carbonaceous condensate, the formation of fullerene cages can be observed. Two cage molecules at the periphery of a denser cluster can be clearly identified in Figure~\ref{fig:HR2}.

Table~\ref{tbl:edx} gives the mean elemental composition of the silicate component of the condensates determined by EDX spectroscopy. The compositions of the silicate grains differ from those of the corresponding targets. An increase of iron and a depletion of magnesium was found for iron-containing silicates, which was caused by an incongruent evaporation of the silicate constituents under laser evaporation. Such discrepancies between the composition of the target and the condensate were already observed in previous experiments. The power density of the laser used for the ablation process of silicate targets was too small to instantly sublimate the atomic and molecular species from the ablation volume. Consequently, a liquid silicate phase was formed that segregated into different phases including a less volatile magnesium-rich silicate and a more volatile iron-rich silicate component. The latter was preferentially evaporated during the laser ablation process.

\begin{deluxetable}{lllll}
\tablecaption{Measured Mean Elemental Composition of the Silicate Component of the Condensates\tablenotemark{a}\label{tbl:edx}}
\tablehead{\colhead{Experiment\tablenotemark{b}} & \colhead{Mg} & \colhead{Fe} & \colhead{Si} & \colhead{O} }
\startdata
E1 &      &      & 47.3 & 52.7 \\
E2 & 27.9 &      & 15.5 & 56.6 \\
E3 &  4.1 & 20.3 & 21.7 & 53.9 \\
E4 &  8.8 & 14.0 & 22.7 & 54.5 \\
\enddata
\tablenotetext{a}{In atomic percent.}
\tablenotetext{b}{Experiments E1, E2, E3, and E4 used targets of SiO, Mg$_2$SiO$_4$, Mg$_{0.4}$Fe$_{0.6}$SiO$_3$, and Mg$_{0.6}$Fe$_{0.4}$SiO$_3$, respectively, along graphite.}
\end{deluxetable}

In addition, the composition of the condensed silicates is not stoichiometric for most of the samples and depends on the composition of molecular species in the ice layer. The formation routes of silicates and carbonaceous solids are highly complex. It is impossible to provide a complete set of reactions leading to refractory siliceous and carbonaceous material simultaneously. For these formation routes, more than a few hundred individual reactions have to be considered. Moreover, the co-condensation of silicates and carbonaceous molecules can result in redox reactions taking place either between oxydizing and reducing molecular species or between reducing and oxidizing solid components. For example, C atoms or small carbon clusters can be oxidized by oxygen-bearing molecules such as SiO, SiO$_2$, Si$_2$O$_3$, FeO, OH, H$_2$O to CO and CO$_2$. Redox reactions additionally complicate the reaction schemes responsible for the formation of both silicates and carbonaceous material.

Solid carbonaceous material can also react as a reducing agent in contact with silicates. The oxidation of carbonaceous material requires the reduction of metallic silicate components. The solid carbon can be oxidized either completely leading to the formation of CO or CO$_2$ or by the formation of oxygen-bearing functional groups such as C$\sbond$OH or C$\dbond$O on the surface. Simultaneously, other components such as Fe cations must be reduced to metallic iron, which is able to form Fe nanometer-sized particles within the silicate matrix.

\begin{figure*}
\epsscale{1.1}
\plotone{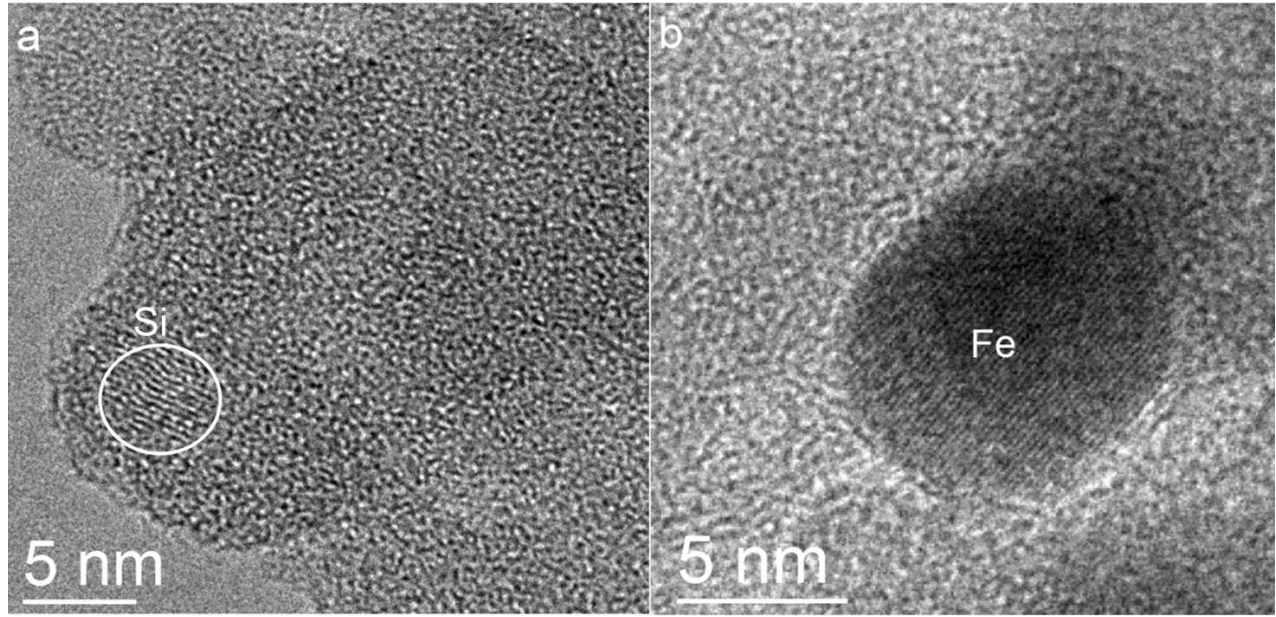}
\caption{(a) HRTEM micrograph of condensed SiO$_x$ grains in experiment E1 with amorphous structure and a small Si inclusion. (b) Metallic iron inclusion formed during the condensation of iron-containing silicate and carbonaceous material in experiment E3.\label{fig:HR3}}
\end{figure*}

Although redox reactions are well known in high-temperature processes, they have not been investigated at cryogenic temperatures yet. The co-condensation of SiO$_x$ and carbonaceous material was found to occasionally produce nanometer-sized silicon crystals clearly visible in Figure~\ref{fig:HR3}. Here, the carbon was oxidized, whereas Si$^{4+}$ was reduced to metallic silicon and could form small crystalline Si inclusions.

The simultaneous condensation of Mg-Fe-silicates with carbonaceous species lead to the reduction of ferric or ferrous ions and the formation of metallic Fe particles. The formation of nanometer-sized silicon crystals and metallic iron inclusions in the carbon-silicate condensates is well documented in Figure~\ref{fig:HR3}. Since Fe$^{2+}$ and Fe$^{3+}$ ions are easier to reduce than Si$^{4+}$ ions, metallic iron particles were frequently produced in these condensation processes.

The HRTEM/EDX analysis does not permit an exact quantification of the silicate/carbon ratio. Nevertheless, a slightly reduced carbon content in the condensate can be derived from the HRTEM images.

One important question is the efficiency of the low-temperature condensation of a refractory material from molecules. It can be roughly determined by comparing the total masses of all atoms and molecules observed in the Ne ice matrix with the mass of the final condensate. Based on a ratio of 0.6~mol magnesium silicate and 0.4~mol carbon grains, an average molecular weight and density was calculated. For the experiment E2 (Mg$_2$SiO$_4$ and graphite targets), a total mass of refractory atoms and molecules, which are available for the condensation process, of 15.3~$\mu$g was calculated using the average concentration of molecules provided in Table~\ref{tbl:concentrations}, the corresponding ice thickness, and the area of the substrate. The amount of solid condensed material was determined using an average thickness of the condensate layers of about 100~$\mu$m determined from scanning electron microscope images, assuming a 90{\%} porosity of the condensate from the electron microscope images and a coverage of the substrate of less than 90{\%} due to the mesh-like topology of the condensate film \citep{Fulvio17}. A dust mass of 14.4~$\mu$g was calculated resulting in a condensation efficiency of 94{\%}.

\section{DISCUSSION}

The debate on the possible growth of dust grains in the ISM has arisen from the discrepancy between the time-scales estimated for injecting the grains of stellar origin into the ISM and for destroying them there. There are actually indications that the dominant source of dust in the ISM is the accretion of gas-phase species onto existing dust grains \citep[][and references therein]{Ginolfi18}.

For instance, the formation and growth of refractory dust grains in the ISM is in agreement with the spatial variations observed in elemental depletion \citep{Draine79,Tielens98,Turner00,Draine09,Jenkins09,Whittet10}, and also with those observed in the Galactic extinction curve \citep{Hirashita12,Hirashita14}. The depletion of iron, which is thought to be the consequence of accretion onto grains in the ISM \citep{Dwek16}, is another indication of interstellar grain growth. Observations of galaxies at early times of the Universe have revealed dust masses that could not be explained if considering dust production by asymptotic giant branch (AGB) stars and supernovae (SNe) only, given the destruction caused by the shocks induced by the very same SNe \citep{Michalowski15}.

Whether or not refractory dust mass is created by accretion of gas-phase species in the ISM, carbonaceous matter and silicates are found as mostly pure grains. Spectropolarimetric measurements of the mid-IR absorption bands that characterize the two materials \citep{Adamson99,Chiar06,Mason07,Li14} suggest their separation. The separation is also supported by the analysis of X-ray halos generated by the scattering of these energetic photons by the dust grains \citep{Dwek04,Smith06,Valencic15}. Analyses of the latter type, however, suggest also grain populations that include, in addition to pure grains, a fraction ($\sim$36\% in mass) of composite grains containing together silicate, refractory organic matter, and water ice \citep{Jin17,Jin19}.

The relevance and results of our experiments with Ne ice matrices are examined with regard to cold interstellar conditions, in which dust formation may proceed according to the following scenarios.

In the conditions of the cold H~{\small I} ($T$ $\sim$100~K, $n_\mathrm{H}$ = 30~cm$^{-3}$) and diffuse H$_2$ ($T$ $\sim$50~K, $n_\mathrm{H}$ $\sim$100~cm$^{-3}$) media, existing grains are bare, hence free atoms and molecules impinge on refractory surfaces. The surfaces are constituted of silicate, carbonaceous, or refractory organic matter. In a hypothetical grain growth scenario, atoms or molecules that stick on such surfaces can diffuse and react with an adequate site and become a component of the refractory matter. The interstellar UV radiation field might maintain the reactivity of the surface of the grains. The accretion process can be facilitated by Coulomb interaction as small grains tend to be neutral or negatively charged whereas gas-phase atoms are more likely neutral or positively charged, including refractory Si, Fe, Mg, C, and O \citep[][and references therein]{Zhukovska16}. On the other hand, electron-tunneling from negatively charged grains to incident cations can reduce the sticking probability \citep{Turner91}. While the chemically compatible species or precursors are incorporated, any other adsorbed substance is removed by sputtering, photodesorption, photolysis, or chemical reaction \citep{Barlow78,Draine09,Jenkins09}. For instance, the growth of amorphous silicate by accretion requires a mechanism that removes C atoms efficiently because they are much more abundant than Si atoms and SiO molecules. It could consist of the formation and desorption of small C-containing molecules, e.g., CH, since atomic hydrogen is orders of magnitude more abundant than any other species. Such mechanism is expected in the case of oxide grains \citep{Duley79,Denison81}. The existence and efficiency of the processes depend on the regional conditions, the formation of a carbon mantle becoming possible as the density conditions shift progressively from diffuse to dense \citep[][and references therein]{Jones17}.

In the dense ISM ($T$ = 10--50~K, $n_\mathrm{H}$ = 10$^3$--10$^6$~cm$^{-3}$), dust grains are covered in ices, composed essentially of H$_2$O, CO, and CO$_2$ molecules \citep[e.g.,][]{Gibb04,Boogert15}, in which the precursors of silicates and carbonaceous matter are embedded. The observation of ions in interstellar ices indicates the existence of bulk diffusion despite the low temperature \citep[][and references therein]{Cuppen17}. Species diffusing in the ice mantle can either reach the surface of the refractory core and become incorporated in case of chemical compatibility, or they can start to nucleate and form new refractory grains in the ice. The latter is the most probable scenario, but it depends on the morphology of the interstellar dust grains and the thickness as well as the structure of the ice. Ice chemistry is complex, however. Irradiation causes the chemical erosion of carbon covered with H$_2$O ice, producing CO and CO$_2$ molecules \citep{Fulvio12,Sabri15}. Thus it hinders the growth and the formation of carbon grains. Irradiation also induces the formation of complex organic molecules like aldehydes, sugars, carboxylic acids, and amino acids \citep[][and references therein]{Oeberg16}. When grains enter intercloud regions, ices are sublimated yet complex molecules remain as an organic refractory residue. The composite grains possibly revealed by some analyses of X-ray halos \citep{Jin17,Jin19} could emerge at that stage. While further irradiation converts the residue into carbonaceous material \citep{Jenniskens93}, grain-grain shocks separate the initial refractory core, which has possibly grown, and the solid nuclei formed in the ice.

In our experiments, the Ne ice matrix can be seen either as the surface of a virtual refractory grain onto which the matrix-isolated atoms and molecules diffuse, or as the ice layer that covers grains in dense interstellar regions. In both scenarios, it acts as the energy sink that a bare grain or an ice mantle constitute for an exothermic reaction. Only in this sense does the Ne matrix simulates the surface of a grain or bulk ice since it does not possess the catalytic or morphological properties of organic ices and refractory surfaces.

With relation to bare refractory surfaces as found in the cold neutral ISM and diffuse ISM, our experiments show that chemical reactions proceed at cryogenic temperatures between SiO molecules, Mg and Fe atoms, or their oxides, and possibly SiO$_2$, to produce complex silicates. Simultaneously, carbon molecules assemble and combine into a solid carbonaceous matter. Still, the precursors of silicates do not react with those of carbonaceous matter. In the absence of selection mechanisms such as photodesorption or sputtering, silicate and carbon materials can stick to each other without combining chemically, as observed in our experiments. Assuming the Ne matrix can be likened to a refractory grain surface and its dopants to reactive sites, our experiments might support the hypothesis of grain growth in the cold neutral ISM and diffuse ISM, silicates and carbonaceous matters growing separately. While it would be favored by Coulomb interaction \citep{Zhukovska16}, the efficiency of the mechanism in its competition against destruction processes must still be evaluated.

Comparing the Ne ice matrix with the ice layer that covers grains in dense interstellar regions, the experiments show that diffusing precursors eventually nucleate to form solid grains of complex silicates and carbonaceous matter at cryogenic temperature despite the presence of H$_2$O and CO molecules, and also CO$_2$. While the formation of carbonates is observed in our experiments, in interstellar ices they would be destroyed by UV irradiation \citep{Ciaravella18}. Since the condensates show silicate and carbon materials stuck to each other, one material may facilitate the condensation of the other, suggesting a possible interfacing of carbon and silicate materials. This is supported by the finding that reduced silicon and iron particles were detected in the complex silicates. The reduction process was triggered by reactions of silicate intermediates, such as (SiO)$_x$ oligomers or small grains, with carbon or CO, which lead to the reduction of Si$^{4+}$, Fe$^{2+}$, and Fe$^{3+}$ and the oxidation of carbon (see Section~\ref{sec:TEM}). The results of our experiments with Ne ice support the notion of grain formation in interstellar ices. Nucleation of precursors is possible since diffusion can be observed in water ice \citep[e.g.,][]{Mispelaer13} and is expected in interstellar ices \citep{Cuppen17}.

\section{Conclusions}
Matrices of Ne ice have been doped, simultaneously, with atoms and molecules that are potential precursors of complex silicates (Mg, Fe, SiO, SiO$_2$) and of carbonaceous materials (C$_n$, $n$ = 2--10). They also contained ice-forming species (CO, CO$_2$, C$_3$O, and H$_2$O) as well as O$_3$. The annealing of the matrices showed the disappearance of the molecular bands and the rise of a broad feature at $\sim$10~$\mu$m in IR spectra measured at temperatures lower than 13~K, thus indicating the formation of amorphous silicate condensates at cryogenic temperatures. This was confirmed by ex-situ HRTEM and EDX spectroscopy analysis. Electron microscopy also revealed the condensation of amorphous carbon in parallel with that of the silicate material. Amorphous carbon was not detected with IR spectroscopy because it does not give rise to defined bands and the amount of material was too little for a noticeable effect on the baseline of the spectra. As a secondary result, the formation of carbonates was also observed, which we attribute to ices rich in carbon oxide molecules in addition to silicate precursors.

Thus both silicate and carbonaceous materials can condense from cold precursors in the absence of radiations similar to interstellar UV photons and cosmic rays. This supports the hypothesis that dust grains are re-formed or grow in the ISM. Moreover, the present observations suggest that species from one of two groups that consist respectively of silicate precursors and carbonaceous matter precursors do not react at cryogenic temperatures with those belonging to the other group. Such finding constitutes a clue as to the separation between silicate and carbonaceous materials observed by astronomers. It is also consistent with the low abundance of SiC grains in the diffuse ISM. On the other hand, silicate precursors may react with carbon oxides in interstellar ices to produce carbonates. Although crystalline carbonates were found in dust shells of evolved stars \citep{Kemper02}, interstellar amorphous carbonates have not been identified.

The results do not allow us to determine whether the formation and/or the growth of dust grains take place exclusively or even preferentially in the cold neutral ISM or in the diffuse ISM through an ice-free process, or in dense clouds where ices would play a role. Experiments with bare and ice-covered refractory surfaces exposed to UV photons are required for this purpose.

\acknowledgments
The authors acknowledge the support of the Deutsche Forschungsgemeinschaft through project No. JA 2107/2-2 within the framework of the Priority Program 1573 "Physics of the Insterstellar Medium". They are most grateful to the anonymous reviewers for comments and suggestions that helped us to considerably improve our manuscript.

\bibliographystyle{aasjournal}
\bibliography{roui1905-rev2}

\appendix

\restartappendixnumbering

\section{Tables}

\startlongtable
\begin{deluxetable*}{llllll}
\tablecaption{Wavelengths of Absorptions Observed in the UV/vis Spectra of Doped Ne Matrices\tablenotemark{a}\label{tbl:UVvisspecies}}
\tablehead{\colhead{Species} & \colhead{E1\tablenotemark{b}} & \colhead{E2} & \colhead{E3} & \colhead{E4} & \colhead{References} }
\startdata
Fe~{\footnotesize I} &    &       & 377.0 & 377.0 &   \\
                     &    &       & 362.6 & 362.6 &   \\
                     &    &       & 335.2 & 335.0 &   \\
                     &    &       & 295.2 & 295.4 &   \\
                     &    &       & 292.4 & 292.4 &   \\
                     &    &       & 288.6 & 289.0 &   \\
                     &    &       & 266.8 & 266.8 &   \\
                     &    &       & 244.2 & 244.0 &   \\
                     &    &       & 238.6 & 239.0 &   \\
                     &    &       & 223.6 & 224.0 &   \\
                     &    &       & 212.6 & 212.4 &   \\
Mg~{\footnotesize I} &    &       & 277.0 & 277.4 &   \\
                     &    & 273.2 & 273.2 & 273.2 &   \\
SiO $A ^1\Pi \leftarrow X ^1\Sigma^+$ & 229.8\tablenotemark{c} &  &  &  & 1 \\
                                      & 225.4                  &  &  &  & 1 \\
                                      & 221.4                  &  &  &  & 1 \\
                                      & 217.4                  &  &  &  & 1 \\
                                      & 213.8                  &  &  &  & 1 \\
                                      & 210.0                  &  &  &  & 1 \\
C$_2$ $D ^1\Sigma_u^+ \leftarrow X ^1\Sigma_g^+$   &       & 232.2 & 232.2 & 232.2 & 2 \\
C$_3$ $^1\Pi_u \leftarrow X ^1\Sigma_g^+$          & 405.6\tablenotemark{d} & 405.6               & 405.6 & 405.6 & 2 \\
C$_6$ (1) $^3\Sigma_u^- \leftarrow X ^3\Sigma_g^-$ & 511.0                  & ns\tablenotemark{e} & ns    & 510.8 & 3 \\
                                                   & 495.4                  & ns                  & ns    &       & 3 \\
                                                   & 473.4                  & ns                  & ns    &       & 3 \\
                                                   & 461.6                  & ns                  & ns    & 461.6 & 3 \\
                                                   & 449.0                  &                     & 448.8 & 449.4 & 3 \\
C$_6$ $^3\Pi_u \leftarrow X ^3\Sigma_g^-$          & 249.6                  &                     &       & 249.6 & 4 \\
                                                   & 245.6                  &                     &       &       & 4 \\
                                                   & 242.0                  &                     &       &       & 4 \\
C$_6$          & 235.0 &       &       &       & 5 \\
               & 233.0 &       &       &       & 5 \\
C$_8$          & 565.0 & ns    & ns    & ns    & 3 \\
CNN            & 419.6 & 419.4 & 419.4 & 419.6 & 2 \\
\enddata
\tablenotetext{a}{In units of nm.}
\tablenotetext{b}{Experiments E1, E2, E3, and E4 used targets of SiO, Mg$_2$SiO$_4$, Mg$_{0.4}$Fe$_{0.6}$SiO$_3$, and Mg$_{0.6}$Fe$_{0.4}$SiO$_3$, respectively, along graphite.}
\tablenotetext{c}{First clear band in the $A ^1\Pi \leftarrow X ^1\Sigma^+$ absorption system. The origin at 234.2~nm \citep{Hormes83} is not clearly seen because of its overlapping with bands of C$_6$.}
\tablenotetext{d}{Head of the $^1\Pi_u \leftarrow X ^1\Sigma_g^+$ absorption system. The positions of the numerous bands in the system are not given.}
\tablenotetext{e}{Not scanned.}
\tablerefs{(1) \cite{Hormes83}; (2) \cite{Weltner66}; (3) \cite{Freivogel95}; (4) \cite{Grutter99}; (5) \cite{Wakabayashi01,Wakabayashi02}.}
\end{deluxetable*}

\clearpage
\newpage

\begin{deluxetable*}{llllllllllll}
\tablecaption{Observed Vibrational Bands of Identified Matrix-isolated Species\tablenotemark{a}\label{tbl:obsvib}}
\tablehead{\colhead{Species} & \colhead{E1} & \colhead{E2} & \colhead{E3} & \colhead{E4} & \colhead{Reference} & \colhead{Species} & \colhead{E1} & \colhead{E2} & \colhead{E3} & \colhead{E4} & \colhead{Reference} }
\startdata
SiO            & 1228.5   & 1228.0   & 1228.0   & 1228.0   & 1, 2\tablenotemark{b}                          & H$_2$O         & 1595.9   &          &          &          & 16 \\
SiO$_2$        &          & 1423.8   & 1423.8   & 1423.8   & This work                                      &                & 1614.2   & 1614.2   & 1614.7   & 1614.2   & 16 \\
SiO$_2$$\sbond$O$_2$ &          &          & 1372.7   & 1372.7   & 3\tablenotemark{c}                             &                & 1630.6   & 1630.6   & 1631.1   & 1631.1   & 16 \\
               &          &          & 1398.7   & 1398.7   & 3\tablenotemark{c}                             &		             & 1649.4   & 1649.4   & 1649.9   & 1649.4   & 16 \\
Si$_2$O$_2$    &  771.4   &          &  771.4   &  771.9   & 1, 2\tablenotemark{b}, 4\tablenotemark{b}      &         			 & 3193.7   &          & 3194.2   &          & 16 \\
               &  809.0   &          &  809.5   &  809.0   & 1, 2\tablenotemark{b}, 4\tablenotemark{b}      & 	        		 & 3680.2   & 3680.7   &          &          & 16 \\
Si$_3$O$_3$    &  977.3   &          &  976.8   &  977.8   & 1, 2\tablenotemark{b}                          &           	 	 & 3696.6   &          & 3697.1   & 3696.6   & 16 \\
C$_3$          & 2036.6   & 2036.6   & 2036.6   & 2036.6   & 5                                              &     					 & 3735.6   & 3735.6   & 3735.6   & 3735.6   & 16 \\
C$_4$          & 1547.1   & 1547.7   & 1547.2   & 1547.2   & 5                                              &                & 3782.9   & 3782.9   & 3782.9   & 3783.4   & 16 \\
C$_5$          & 1444.0   & 1444.5   & 1444.5   & 1444.5   & 5                                              &         			 & 3801.7   & 3802.6   & 3804.6   & 3804.1   & 16 \\
               & 2166.7   & 2166.7   & 2166.7   & 2166.7   & 5                                              &                & 5359.0   & 5359.0   & 5359.0   & 5359.0   & 16 \\
C$_6$          & 1199.6   &          & 1199.6   & 1199.6   & 5                                              & (H$_2$O)$_2$   & 1598.8   & 1598.8   & 1598.8   & 1599.3   & 16 \\
               & 1958.9   & 1958.9   & 1958.9   & 1958.9   & 5                                              & 							 & 1616.6   & 1616.6   & 1616.6   & 1616.6   & 17 \\
C$_7$          & 1897.7   & 1897.7   & 1897.7   & 1897.7   & 5, 6                                           &                & 1882.8   &          &          &          & 18 \\
               & 2135.9   & 2135.9   & 2135.9   & 2135.9   & 5, 6                                           &                & 3590.0   & 3590.0   & 3590.0   & 3590.0   & 17, 18 \\
C$_8$          & 1707.7   & 1708.2   & 1707.7   & 1707.7   & 7                                              & 							 & 3659.5   & 3659.5   & 3659.5   & 3659.5   & 17, 18 \\
               & 2067.9   & 2067.9   & 2067.9   & 2067.9   & 7, 8                                           &                & 3733.2   & 3733.2   & 3732.7   & 3733.2   & 17, 18 \\
C$_9$          & 2010.5   & 2010.5   & 2010.5   & 2010.5   & 5                                              & 							 & 3763.6   & 3762.1   & 3763.6   & 3763.6   & 17, 18 \\
               & 2081.4   &          & 2081.9   & 2081.4   & 9                                              & (H$_2$O)$_3$   & 1607.9   &          & 1607.9   & 1607.5   & 19 \\
C$_{10}$       & 2075.6   & 2075.6   &          & 2075.6   & 7                                              &                & 3529.3   & 3527.8   & 3529.8   & 3530.2   & 16, 19 \\
CO             & 2141.2   & 2141.2   & 2141.2   & 2141.2   & 10                                             & (H$_2$O)$_4$   & 3381.3   & $\sim$3375 & $\sim$3381 &        & 20 \\
CO$_2$         &          &  667.8   &  668.2   &  668.2   & 11                                             & CO$\sbond$H$_2$O		 &          & 2151.8   & 2151.8   &          & 10 \\
               & 2348.0   & 2347.5   & 2347.5   & 2347.5   & 11                                             & O$_3$          &          & 1040.0   & 1040.0   & 1040.0   & 21 \\
               &          & 3611.7   & 3611.7   & 3611.7   & 11                                             & O$_4^+$        &          & 1164.4   & 1164.9   & 1164.9   & 22, 23 \\
               &          & 3713.9   & 3713.9   & 3713.9   & 11                                             & FeO            &          &          &  873.2   &          & 24\tablenotemark{c} \\
$^{13}$CO$_2$  &          & 2282.0   & 2282.0   & 2282.0   & 11                                             & FeOFe          &          &          &  869.8   &  869.8   & 25\tablenotemark{c} \\
CO$^{18}$O     & 2330.7   & 2330.7   & 2330.7   & 2330.7   & 11                                             & OFeO           &          &          &  950.3   &  950.3   & 25\tablenotemark{c} \\
C$_2$O         &          & 1972.9   & 1972.9   & 1972.9   & 12                                             & Fe(O$_2$)      &          &          &  957.3   &          & 25\tablenotemark{c} \\
C$_3$O         & 2253.5   & 2253.5   & 2253.5   & 2253.5   & 13\tablenotemark{c},14\tablenotemark{d},15\tablenotemark{c} & & & & & & \\
\enddata
\tablenotetext{a}{In units of cm$^{-1}$. Unless otherwise indicated, references concern measurements in Ne matrix.}
\tablenotetext{b}{In N$_2$ matrix.}
\tablenotetext{c}{In Ar matrix.}
\tablenotetext{d}{In gas phase.}
\tablerefs{(1) \cite{Hastie69}; (2) \cite{Khanna81}; (3) \cite{Schnoeckel78}; (4) \cite{Anderson68}; (5) \cite{Smith94}; (6) \cite{Forney96}; (7) \cite{Freivogel97}; (8) \cite{Freivogel95}; (9) \cite{Szczepanski96}; (10) \cite{Dubost76}; (11) \cite{Wan09}; (12) \cite{Fulara98}; (13) \cite{Botschwina91}; (14) \cite{McNaughton91}; (15) \cite{Strelnikov04}; (16) \cite{Forney93}; (17) \cite{Ceponkus08}; (18) \cite{Ceponkus04}; (19) \cite{Ceponkus05}; (20) \cite{Ceponkus12}; (21) \cite{Brosset93}; (22) \cite{Thompson89}; (23) \cite{Jacox94}; (24) \cite{Green79}; (25) \cite{Chertihin96}, whose assignment of FeOFe is not supported by the theoretical calculations of \cite{Gutsev03} and ours.}
\end{deluxetable*}

\end{document}